\documentclass[twocolumn]{aastex631}
\usepackage{amsmath,amssymb,amstext,bm}
\usepackage{graphicx, enumerate}
\usepackage{wasysym}
\usepackage{xcolor}
\usepackage{soul}

\newcommand{\kms}{\ensuremath{\rm km\,s^{-1}\,}}
\usepackage{blindtext}
\usepackage{ulem}
\usepackage{chngcntr} 

\shorttitle{outflow knots driven by multiple velocity modes}
\shortauthors{Lora et al.}

\begin{document}
\title{Shedding light on the ejection history of molecular outflows:\\
Multiple velocity modes and precession}

\author[0000-0003-3588-5235]{V. Lora}
\affiliation{Instituto de Ciencias Nucleares, Universidad Nacional Aut\'onoma de M\'exico,\\
Apartado Postal 70-543, 04510 Ciudad de M\'{e}xico, Mexico}

\author[0000-0003-3246-0821]{T. Nony}
\affiliation{Instituto de Radioastronom{\'i}a y Astrof{\'i}sica, Universidad Nacional Aut\'onoma de M\'exico,\\
Apartado Postal 72-3 (Xangari), Morelia, Michoc{\'a}n 58089, Mexico}
\affiliation{INAF, Osservatorio Astrofisico di Arcetri, 
Largo Enrico Fermi, 5, I-50125 Firenze, Italy}

\author[0000-0001-7222-1492]{A. Esquivel}
\affiliation{Instituto de Ciencias Nucleares, Universidad Nacional Aut\'onoma de M\'exico,\\
Apartado Postal 70-543, 04510 Ciudad de M\'{e}xico, Mexico}

\author[0000-0003-1480-4643]{R. Galv\'an-Madrid}
\affiliation{Instituto de Radioastronom{\'i}a y Astrof{\'i}sica, Universidad Nacional Aut\'onoma de M\'exico,\\
Apartado Postal 72-3 (Xangari), Morelia, Michoc{\'a}n 58089, Mexico}

\begin{abstract}
Variable accretion has been well studied in evolved stages of low-mass stars formation. However, the accretion history in the initial phases of star formation is still a seldom studied topic. The outflows and jets emerging from protostellar objects could shed some light onto their accretion history.
We consider the recently studied case of W43-MM1, 
a protocluster containing 46 outflows driven by 27 protostellar cores. The outflow kinematics of the individual cores and associated knots in W43-MM1 indicate episodic protostellar ejection. We took the observed parameters of an individual core system (core \#8) and performed 3D hydrodynamic simulations of such system including  episodic changes in the velocity of the outflow.
The simulations consist of a collimated jet emerging from a 
core taking into account one- and two-velocity modes in the variation of the ejection velocity of the jet. In addition, we investigated the effect of including the precession of the jet to the one- and two-velocity mode models. From the simulations, we constructed position-velocity diagrams and compared them with the observations. We find that including a second mode in the ejection velocity,  as well as the precession, are required to explain the positions of the outflow knots  and other position-velocity features observed in the outflow of core \#8 in W43-MM1.

\end{abstract}


\section{Introduction}\label{sec:intro}
 
The time variability of gas accretion during star formation is an important and debated topic. While low-mass, pre-main sequence stars have been statistically studied on short time scales, the variability of accretion during the embedded protostellar phase is not well documented \citep[see, e.g., reviews by][]{Audard14,Fisher22}.
Yet, it is during this phase that most of the final stellar mass accumulates from the gas reservoir. 
Observations of few protostars have brought evidence of quasi-periodic ligh curves, interpreted as accretion variability \citep[see e.g. V2492 Cyg][]{Hillenbrand13}.

Jets and outflows develop as protostellar accretion occurs and they aid in the removal of angular momentum from the collapsing protostellar core. The morphology and kinematics of molecular outflows could thus provide fossil records of the accretion history of a protostar \citep{Rohde22}.
Molecular outflows from young protostars extend up to a few $\sim$0.1~pc, with collimation angles of a few degrees only, and velocities up to $100~\kms$ \citep[see, e.g., the review by][]{Bally16}. They are often composed of a collimated jet at high velocity, plus a shell-like wind tracing gas at lower speeds. Collimated jets frequently appear as chains of knots interpreted as internal shocks produced by episodic variations in the protostar mass-loss rate and/or the ejection velocity.

Position-velocity (PV) diagrams of these jets often display velocity spurs referred to as ``Hubble laws'', since the gas velocity of these ejection phenomena increases linearly with the distance to the protostar \citep[e.g., the case of the class 0 object HH211,][]{Gueth99}. A series of these velocity spurs composes a jagged PV profile sometimes called ``Hubble wedge'' \citep{Arce01a}. These complex PV diagrams have been investigated in about a dozen of low-mass Class~0 protostars so far,  including HH111 \citep{Cernicharo96}, L1551 \citep{Bachiller94}, L1157 \citep{Gueth98}, HH212 \citep{Lee06}, HH46-47 \citep{Arce13}, and more recently Cep. E \citep{Schutzer22}. 

In the massive protocluster W43-MM1, \cite{Nony20} discovered a cluster of 46 outflows driven by 27 protostellar cores, with masses in the range from $1$ to $100$ M$_{\odot}$. Their detailed investigation of outflow kinematics using PV diagrams reveled clear events of episodic ejection. A typical ejection timescale (period) of about 500 yr has been estimated from the positions of 86 knots along 38 outflow lobes. 

Based on the observational evidence, \citet{1990ApJ...364..601R} proposed to include an intrinsic variability in the source of jets to model stellar outflows, and was soon applied to specific objects, as the case of HH34 \citep{1992ApJ...386..222R}. An analytical description of the formation of the shocks using the viscous Burgers equation was given in \citet{1992ApJ...390..359K}. The solution of the resulting  ``internal working surfaces'' for a sinusoidal variation can be found in \citet{1998RMxAA..34...73R}.
This model has been widely used to explain Herbig-Haro (HH) objets, for instance, \citet{castellanos:18} have modeled the optical HH1 jet as the result of a two-mode periodic velocity  jet. The model reproduces remarkably well the chain of knots seen close to the source, as well as the time evolution  which is resolved by multi-epoch HST observations.

The motivation to include sinusoidal time dependencies in the ejection velocities is therefore phenomenological. The physical origin of the velocity variability is still unclear, but different explanations have been proposed. For example, the variability could be due to disk thermal instabilities caused by a runaway process. If the temperature in the accretion disk increases to reach the hydrogen ionization temperature, the 
opacity in the disk increases as $\kappa \propto T^{10}$ \citep{Audard14}
trapping the energy inside the disk and causing a rise in the temperature (thermal runaway). The viscosity is proportional to the temperature in the disk, and to the accretion rate, therefore an enhancement in temperature causes an increase in the accretion rate. Another way of inducing a thermal instability could be due to the presence of a companion \citep{Lodato04}, or thermal instability could be triggered by the combination of gravitational and magneto rotational instabilities \citep{Martin2011, Martin2013}. Moreover, disk fragmentation could also cause mass accretion and luminosity outbursts, when the material migrate inwards onto the  protostar.

The different types of instabilities mentioned before, could cause quasi-periodic variations in the accretion and ejection phenomena on timescales as long as 50 kyr \citep{Audard14}. It would be expected that these enhancements are stronger than the underlying stochastic variability, which could also be present. 
Any particular form of periodic variation in a supersonic jet would form internal shocks due to the steepening of the wave-form, and in the hyper-sonic limit can be described by the Burgers equation 
\citep{1992ApJ...390..359K, 1998RMxAA..34...73R}. Thus, the choice of periodic sinusoidal variations is a reasonable  empirical description in jets and outflows with an observed periodicity in their knots.

Only few models have been compared with observations of molecular outflows using PV diagrams, and they strongly depend on the structure adopted for the protostellar envelope \citep[e.g.,][]{Rohde19}. 
Two recent works use PV diagrams to compare their simulations with observations of molecular outflows. \cite{Rabenanahary22} compared simulations of an outflow driven by a pulsed jet and observations in CARMA-7 by \cite{Plunkett15}. \cite{Rivera23} carried out simulations of a jet aimed at reproducing the Cepheus-E outflow observed by \cite{Schutzer22}.

Motivated by the observations of \cite{Nony20}, which show chains of knots arising from dense cores forming intermediate- to high-mass stars, we performed 3D hydrodynamic simulations of the outflows emanating from dense cores. In particular, we explore if a variability in velocity and precession can explain the kinematic structures emerging from the outflow of core MM1\#8 in \cite{Nony20}.

The paper is organized as follows: In Section~\ref{sec:observations} we briefly present the observations of the W43-MM1 protocluser. In Section~\ref{sec:sims} we describe our 3D hydrodynamical numerical simulations. In Section~\ref{sec:results} we report the results of the simulations, and the comparison with the observations. Finally, in Section~\ref{sec:conclusions} we summarize our main conclusions.

\section{W43-MM1 protocluster data}
\label{sec:observations}

This work makes use of ALMA Band 6 observations of the W43-MM1 protocluser. The continuum and lines images have an angular resolution of about 0.5$\arcsec$ ($\sim 2200$\,au at a distance of 5.5 kpc). A first catalogue of cores has been published by \cite{motte18b}, and the association between cores and molecular outflow identified in CO(2--1) has been presented in \cite{Nony20}. 

In the following, we focus on the outflow driven by core \#8. This core has a gas mass of $18\pm3$ M$_{\odot}$ and a systemic velocity of $V_{\rm LSR}=97\,\kms$. 
(see \citealt{Nony20} and \citealt{Nony23}). 
Core \#8 has been chosen for the unique morphology of its bipolar outflow. Indeed, the blue-shifted lobe is the longest of the region, with a total length of 0.44~pc, and therefore the more substructured (eight knots). Its red-shifted counterpart extends over 0.13 pc and has four knots. The relative positions and velocities of the knots reported in this study are taken from Table 3 of \citet{Nony20}, which we include in Table \ref{table:knots}. Their velocity range is from 36$\,\kms$ to 79$\,\kms$ from the core $V_{LSR}$.

\begin{deluxetable}{ccccccc}
    \tablecolumns{7}
    \tablewidth{0pc}
    \tablecaption{Parameters of the blue- and red-shifted knots of core \#8. Taken from Table 3 of \cite{Nony20}, except for B1a (see Section~\ref{sec:pv}).}
    \tablehead{
    Knot & $r_{knot}$ & $\Delta V_{knot}$  & Lobe  \\
    number & ['']     & [km/s]             &   }
    \startdata
    B0  & 0.6  & 47 & Blue \\
    B1  & 1.8  & 66 & Blue \\
    B1a$\star$ & 3.8  & 59  & Blue \\
    B2  & 5.3  & 79  & Blue \\
    B3  & 6.8  & 31  & Blue \\
    B4  & 8.3  & 34  & Blue \\
    B5  & 11.2 & 50  & Blue \\
    B6  & 14.0 & 51  & Blue \\
    B7  & 16.7 & 36  & Blue \\
    R1  & 0.8  & 51  & Red \\
    R2  & 3.0 & 68  & Red \\
    R3  & 3.9 & 48  & Red \\
    R4  & 6.2 & 49  & Red 
    \label{table:knots}
    \enddata
\end{deluxetable}

\begin{figure*}
\centering
\includegraphics[width=0.49\linewidth]{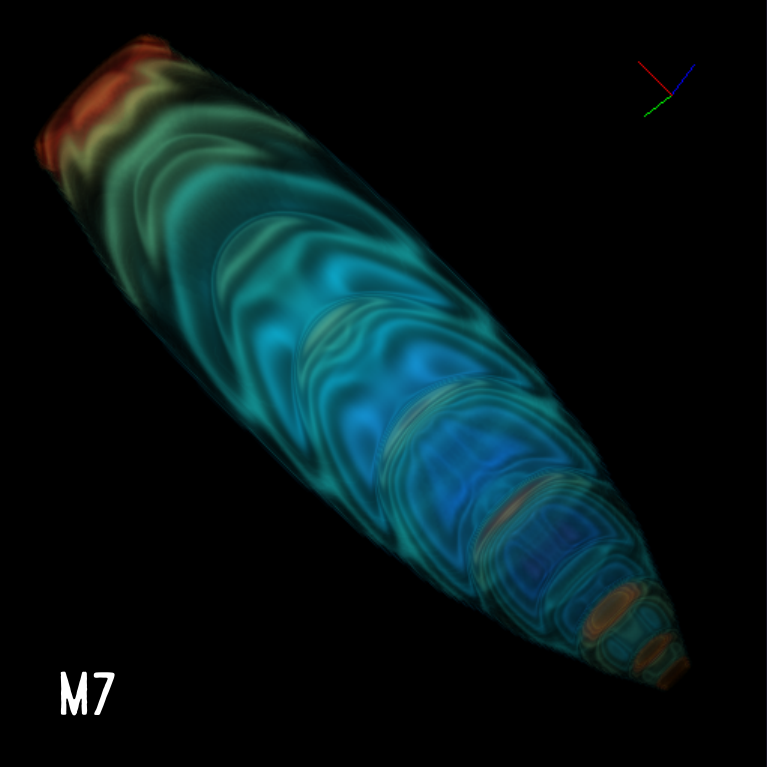}
\includegraphics[width=0.49\linewidth]{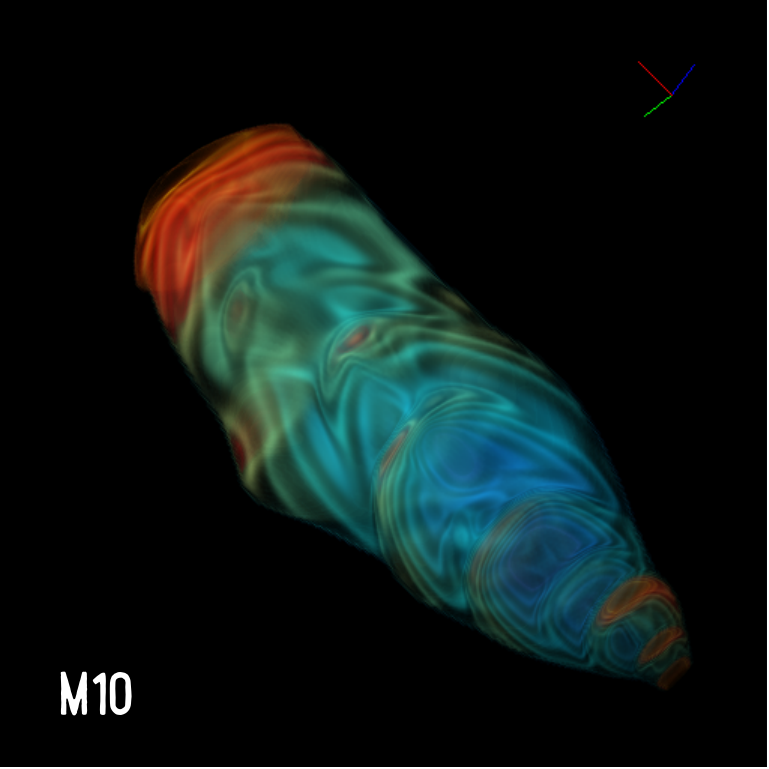}
\caption{Example of two synthetic nebular images (renders) of the jet material of simulations M7 (left panel) and M10 (right panel). See Table \ref{table:simulations} for details on the simulations. 
Blue-green tinges show lower number densities ($10^4$ cm$^{-3}$). Yellow-red tinges show higher number densities ($10^5$-$10^6$ cm$^{-3}$). The positions of the knots (in red) are clearly distinguished at the highest densities.}
\label{fig:renders}
\end{figure*}

\section{Numerical Simulations}\label{sec:sims}
\subsection{The jet model}\label{sec:numerical_model}
A common feature in low-mass protostellar outflows (i.e., HH objects) is a chain of aligned ``knots'' close to the source, which can be produced by a variability in the ejection. The most accepted  model to produce the observed characteristics of this chain of knots is to consider a periodic velocity ejection variation \citep[see for instance][]{1994ApJ...434..221B}. In this scenario, high(er) velocity material catches up with previously ejected material, coalescing into a double shock structure also known as  \emph{internal working surface}. These internal working surfaces are precisely the knots, which move as a unit away from the source. The distance between successive knots is given by the variability of the ejection velocity.

In some cases, such as HH 1 and HH 34 \citep{Raga:11,Raga:12}, besides the chain of knots close to the source, a second series of one or more knots are observed at larger distances. In order to reproduce the morphology of HH 34, \citet{Raga:98,Raga:11,Raga:12} included two or three modes in the ejection velocity variability. More recently, \citet{castellanos:18} implemented a two-velocity variable mode to reproduce the morphology of HH 1. The variabilities in the jet velocity produce a series of two-shock internal working surfaces with maximum shock speeds comparable to the half-amplitude of the ejection velocity variability. 

An alternative to a velocity variation that could produce a chain of knots is to consider a constant velocity and a time-dependent density. However, this scenario does not produce sufficiently strong shocks to account for the excitation state observed in the knots. 
In that case strong shocks are only produced when the material interacts with the environment, but there are no strong shocks in between knots.

\citet{castellanos:18} point out that such variability  could become important if the jet has a very substantial precession, in such a way that the denser regions of the jet can interact with the surrounding environment.

Interestingly, \citet{Nony20} also proposed that the observed changes in the direction of the outflow of core \#8 
could be the consequence of periodic oscillations due to the precession of the molecular jet. The latter work prompted us to study core \#8 in detail using 3D-hydrodynamical simulations of a jet with one- and two-mode velocity variability, which presents (or not) precession. In the following sections we describe the code used (Section \ref{sec:code}), and summarize the set-up of the simulations (Section \ref{sec:setup}).

\begin{figure*}
\epsscale{1.2}
\plotone{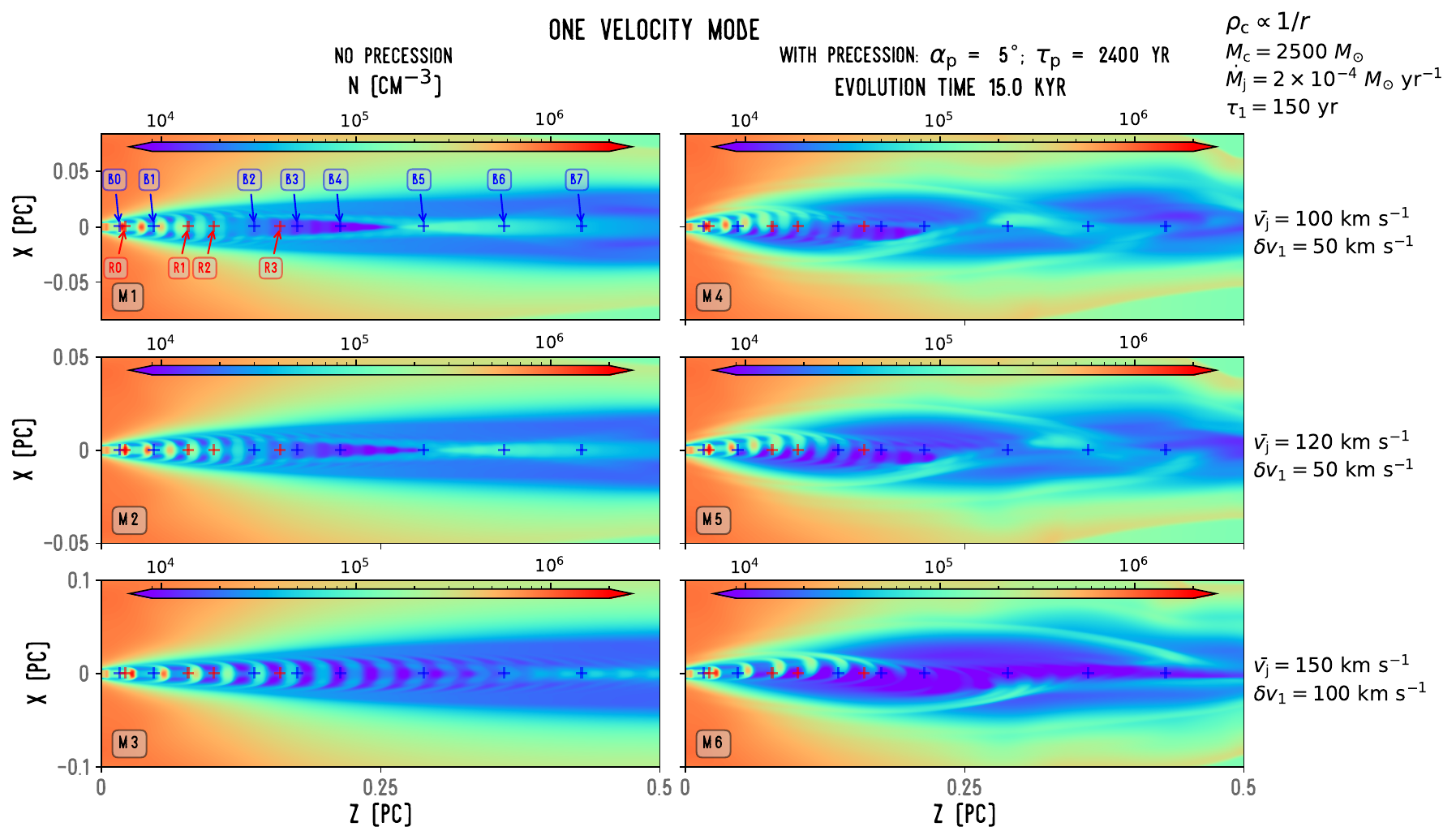}
\caption{Density cuts in the $XZ$-midlplane for the single-velocity mode models. In the left column we present the models without precession, and in the right column the models with precession (with a half opening angle of $5\degr$ and a precession period of $2400~\mathrm{yr}$). In different rows we show models with different velocity variability. The average velocity of the jet and the amplitude of the fluctuation are displayed at the right of each row. Blue and red crosses  mark the position of the blue- and red-shifted knots, respectively, reported by \cite{Nony20} for core \#8 (see Table~\ref{table:knots}).
\label{fig:dens1}}
\end{figure*}

\subsection{The code}\label{sec:code}
To model the interaction of a jet from a core with its environment we use the {\sc guacho} code \citep{2009A&A...507..855E,2013ApJ...779..111E,2018MNRAS.479.3115V}.
The version of the code we used solves the ideal hydrodynamic equations (Equations \ref{ec:1}-\ref{ec:3}) along with a rate equation for hydrogen (Equation \ref{eq:rate}):
\begin{equation}
    \frac{\partial \rho}{\partial t } + \bm{\nabla \cdot }\left(\rho\bm{u} \right) = 0,
\label{ec:1}
\end{equation}

\begin{equation}
    \frac{\partial \left( \rho \bm{u} \right)}{\partial t} + \bm{\nabla \cdot }\left(\rho\bm{u u}\right)+ \nabla P = 0,
\label{ec:2}
\end{equation}

\begin{equation}
    \frac{\partial E}{\partial t} + \bm{\nabla \cdot}\left[\bm{u}\left(E+P\right) \right] = G - L,
\label{ec:3}
\end{equation}

\begin{equation}
\frac{\partial n_\mathrm{HI}}{\partial t} + \nabla (n_\mathrm{HI}\bm{u}) =
    n_\mathrm{e} n_\mathrm{HII} \alpha(T)  -  n_\mathrm{HI} n_\mathrm{e} c(T).
\label{eq:rate} 
\end{equation} 

\begin{figure*}[t]
\epsscale{1.2}
\plotone{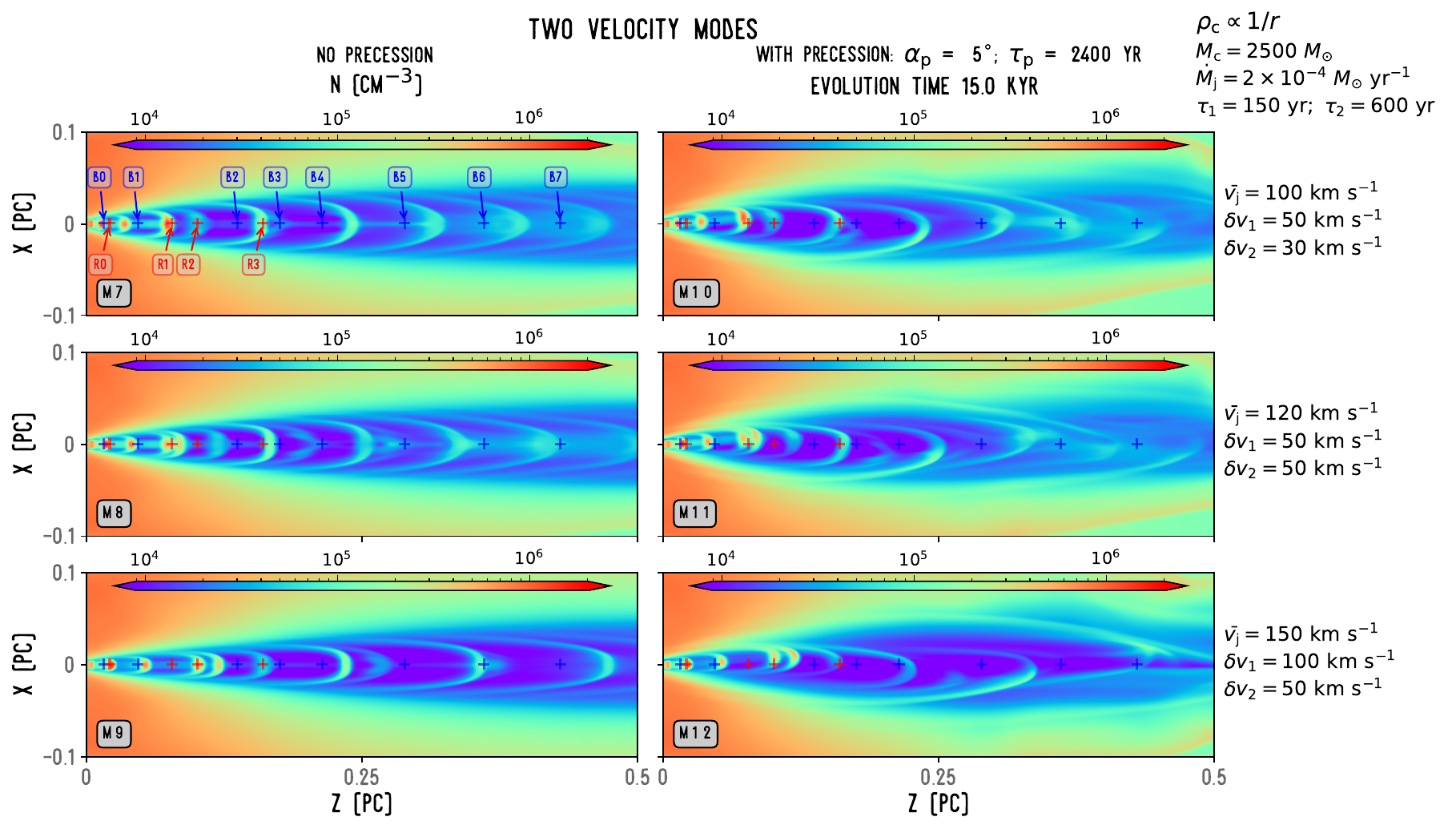}
\caption{Same as Figure \ref{fig:dens1} but for the models with two-velocity modes. The parameters of these are indicated at the right of each row of plots. 
\label{fig:dens2}}
\end{figure*}

In the latter set of equations (Equations \ref{ec:1}-\ref{eq:rate}), $\rho$, $\bm{u}$, and $P$ are the (total) mass density, flow velocity, and thermal pressure. $E$ is the energy density, $G$ and $L$ are the energy gain and losses (respectively) due to radiative processes.
An ideal gas equation of state is used to relate the total energy density with the kinetic energy density and thermal pressure as $E = \rho \vert \bm{u}^2 \vert/2+ P/(\gamma-1)$, where $\gamma=5/3$ is the ratio
between the specific heat capacities. For the hydrogen rate equation (Equation \ref{eq:rate}), $n_\mathrm{HI}$ is the neutral hydrogen (number) density, $n_\mathrm{e}$ the electron density, and $n_\mathrm{HII}$ the ionized hydrogen density. We assume that all the material consists of hydrogen, that all the free electrons arise from ionization of hydrogen ($n_\mathrm{e}=n_\mathrm{HII}$), and neglect the mass of such electrons to compute the mass density, i.e. $\rho \approx m_\mathrm{H}\left( n_\mathrm{HI} + n_\mathrm{HII}  \right)$, where $m_\mathrm{H}$ is the hydrogen mass. In the source terms in Equation (\ref{eq:rate}), $\alpha(T)$ and $c(T)$ are the (case B) radiative recombination coefficient and the collisional ionization coefficient, respectively.
The equations are solved in a regular Cartessian grid with a second order accurate Godunov type method that uses the HLLC approximate Riemann Solver \citep{torobook}, and a linear reconstruction with the $\mathrm{minmod}$ slope limiter to ensure stability.
We do not include radiative gains, but we compute the radiative cooling using the hydrogen ionization fraction following the prescription given by \citet{Biro1995}, and add it to the energy equation with the implementation described in \citet{2021MNRAS.501.4383V}.

In our simulations we do not consider magnetic fields. The magnetic field plays a crucial role in the launching mechanism of the jets \citep{Shu00, Konigl00}, and it is also regarded as important to maintain their collimation at large distances from the source \citep{1982MNRAS.199..883B}. However, our models deal with scales larger than the injection of the jet, and since we impose a collimated flow, it is reasonable to disregard the magnetic effect as a first approximation. This can be further justified considering that the ram pressure in the jets produced in our models is  on the order of $10^{-4}~\mathrm{dyn\,cm^{-3}}$ (e.g. a number density of $\sim10^6~\mathrm{cm^{-3}}$ and a velocity of $\sim 100~\kms$), which is several orders of magnitude larger compared to the magnetic pressure of $\sim4\times 10^{-8}~\mathrm{dyn\,cm^{-3}}$ assuming a typical value of $\vert B\vert =  1~\mathrm{mG}$.

\subsection{Numerical Setup}\label{sec:setup}
We ran a set of 3D-hydrodynamical simulations of a collimated flow arising from a core. 
The extent of the computational domain is $0.17\times0.17\times0.5~\mathrm{pc}$ in the ($x, y, z$) directions, with a uniform resolution of $8.33\times 10^{-4}~\mathrm{pc}$ (i.e. a grid with $200\times200\times600$ cells).

The jet is assumed to travel into a stationary environment with the characteristics of a clump with a (radial) density profile of the form:
\begin{equation}
    \rho_c(r) = 
    \begin{cases}
        \rho_0  & \text{for $r \leq r_0$}, \\
        \rho_0  \left(\frac{r_0}{r}\right) & \text{for $r > r_0$},
    \end{cases}
\end{equation}
\label{eq:rho_c}

where $r_0 = 0.05~\mathrm{pc}$, and the value of $\rho_0$ is set to result in a mass of $2500$ M$_{\odot}$ within a radius of $0.5~\mathrm{pc}$ (yielding $\rho_0\approx 2.1\times 10^{-18}~\mathrm{g~cm^{-3}}$). The temperature of the environment is set to $T_c=1000$ K. Even if this value is high, by running control simulations we found out that the medium temperature does not change our results; the important aspect being the pressure ($nT$) contrast between the medium and the jet.

Therefore, we considered a jet also with a temperature of $T_j = 1000$ K, and a (mean) mass loss rate of $\dot{M}_j=2\times10^{-4}$ M$_{\odot}$ yr$^{-1}$.
The density of the jet is constant, and its value is obtained for each model from the mass loss rate as:
\begin{equation}
    \rho_j = \frac{\dot{M}_j}{4\pi\,v_1\,R_j^2},
    \label{eq:rho_j}
\end{equation}
where $v_1$ is the mean velocity of the jet, and $R_j$ its cylindrical radius.
The jet is imposed at every time-step in a cylindrical region of length $L_j=1.6\times10^{16}\,\mathrm{cm}$, and radius $R_j=1.6\times10^{16}\,\mathrm{cm}$. The injection velocity is along the cylinder symmetry axis.
The interaction between the jet and the dense molecular envelope from the clump results in the formation of a layer of entrained gas which follows the motion of the central jet. The resulting ejected (jet) and entrained material is, at first order, representative of the observed molecular outflow.

For the velocity of the jet, we consider a sinusoidal variability in the ejection velocity. We first considered a one-mode velocity variability mode described of the form: 

\begin{equation}
    v_j(t) = v_1 +\delta v_1 \sin\left( \frac{2\pi t}{\tau_1}\right) \mbox{ .}
    \label{eq:one_mode}
\end{equation}
Secondly, we considered a two-velocity variability mode described by:
\begin{equation}
    v_j(t) = v_1 + \delta v_1 \sin\left(\frac{2\pi t}{\tau_1}\right) 
                 + \delta v_2 \sin\left(\frac{2\pi t}{\tau_2}\right) \mbox{ .}
    \label{ec:two_modes}
\end{equation}
In the models, we explored  three different values of the average velocity of the jet; ($\bar{v_j} = v_1 = 100, 120$, and $150~\kms$). The velocity has a fast high-amplitude mode $\delta v_1$; we explore two different values $\delta v_1 = 50$ and $100$~\kms, with a period of $\tau_1 = 0.15$ kyr.
For the two-mode variability mode in the velocity, a second term is added (see Equation \ref{ec:two_modes}) corresponding to a slow lower-amplitude mode $\delta v_2$; we explored two different values $\delta v_2 = 30$ and $50$~\kms, with a period of $\tau_2 = 0.6$ kyr.

Besides the one- and two-velocity variability modes, we also investigated the effect of the precession of the jet. Therefore, we included models in which the jet precesses with a half-opening angle of $\alpha_p = 5^{\circ}$ and a period of $\tau_p = 2.4$ kyr. In total we performed 12 simulations, the initial conditions and relevant parameters are summarized in Table \ref{table:simulations}.

\begin{deluxetable}{ccccc}
    \tablecolumns{5}
    \tablewidth{0pc}
    \tablecaption{Initial conditions of the simulations; each of the columns correspond to (1) model number, 
    (2) average outflow velocity,
    (3) fast-low-amplitude mode,
    (4) slow-high-amplitude mode,
    and (5) simulation including precession with half opening angle $\alpha_p = 5^{\circ}$ and a precession period of $\tau_p = 2.4$ kyr. All models include a velocity variability with period $\tau_1=150~\mathrm{yr}$, while models 7 through 12 include a second variability mode with period $\tau_2=600~ \mathrm{yr}$. }
    \tablehead{
    Model & $v_1$ & $\delta v_1$ & $\delta v_2$ & precession \\
     & [km/s] & [km/s] & [km/s] &  }
    \startdata
    M1  & 100 & 50 & ---  & No \\
    M2  & 120 & 50 & ---  & No \\
    M3  & 150 & 100& ---  & No \\
    M4  & 100 & 50 & ---  & Yes\\
    M5  & 120 & 50 & ---  & Yes\\
    M6  & 150 & 100& ---  & Yes\\
    M7  & 100 & 50 & 30   & No \\
    M8  & 120 & 50 & 50   & No \\
    M9  & 150 & 100& 50   & No \\
    M10 & 100 & 50 & 30   & Yes \\
    M11 & 120 & 50 & 50   & Yes \\
    M12 & 150 & 100& 50   & Yes
    \label{table:simulations}
    \enddata
\end{deluxetable}

\section{Results}\label{sec:results}
We let the models described in Section \ref{sec:setup} to run for 20~kyr.
As an example, in Figure \ref{fig:renders} we present two density render images of two different models (M7 and M10, see Table \ref{table:simulations}). The transparency in both panels of Figure \ref{fig:renders} vary linearly with density, meaning that denser regions appear more opaque.
Two different tinges represent two different density values; blue-green tinges show lower number densities ($10^4$ cm$^{-3}$), and yellow-red tinges show lower number densities ($10^5$-$10^6$ cm$^{-3}$).
The rendered images were created using the Multi-code Analysis Toolkit for Astrophysical Simulation Data {\sc yt} \citep{Turk2011}. 

\subsection{The one velocity mode case}\label{sec:one_velocity}
In Figure \ref{fig:dens1} we show a density cut in the X-Z midplane of the simulation after an integration time of $15~\mathrm{kyr}$. The associated column density maps are shown in Figure \ref{fig:N1}.
The separation between successive knots in a single velocity mode case is given by  $\Delta x \approx v_1 \tau_1$ which for our models without precession; M1, M2 and M3 (left panels of Figure \ref{fig:dens1}), are $\Delta x1_\mathrm{M1} = 0.015$\,pc, $\Delta x1_\mathrm{M2} = 0.018$\,pc, and $\Delta x1_\mathrm{M3} = 0.022$\,pc, respectively. 

If we compare models M1 and M2, whose only difference is the mean jet velocity (by $20~\kms$), we can see that the chain of knots is evident at larger distances for the model with higher mean velocity (M2, middle row in the left column of Figure \ref{fig:dens1}).
The reason for this is a combination of the higher velocity of the flow itself, with the fact that the leading shock in faster models has the opportunity to clear more material ahead of the source.  Thus, the subsequent internal working surfaces can travel more easily into the low-density canal carved by the leading shock. We can verify this in model M3, which has a higher mean velocity of $150~\kms$.

For visual reference, in Figure \ref{fig:dens1} we over-plotted the positions of the knots of core \#8 measured by \citet{Nony20} and reported in Table \ref{table:knots}.
The one-velocity mode reproduces fairly well the knots closer to the source, but this model fails to reproduce the (blue) knots at farther distances ($>0.187$ pc, B4 to B7). 
The change of direction in the ejected material in the models with precession (right column in Figure  \ref{fig:dens1}) results in some of the material colliding with the ``walls'' of the low-density cavity carved be the leading shock (and in this case also by subsequent internal working surfaces). This oblique collision results in a density increase and a slight deceleration of the knots that collide with the wall. In the case of the one-mode variability models we see a density enhancement at a distance of $\sim 0.31$ pc which can be associated with one of the knots farther away from the source
(B5; top-right panel of Figure \ref{fig:dens1}, and B7; middle-right panel of Figure \ref{fig:dens1}). However, it fails to reproduce the other blue knots observed at greater distances ($>0.31$ pc).

\begin{figure*}
\epsscale{1.2}
\plotone{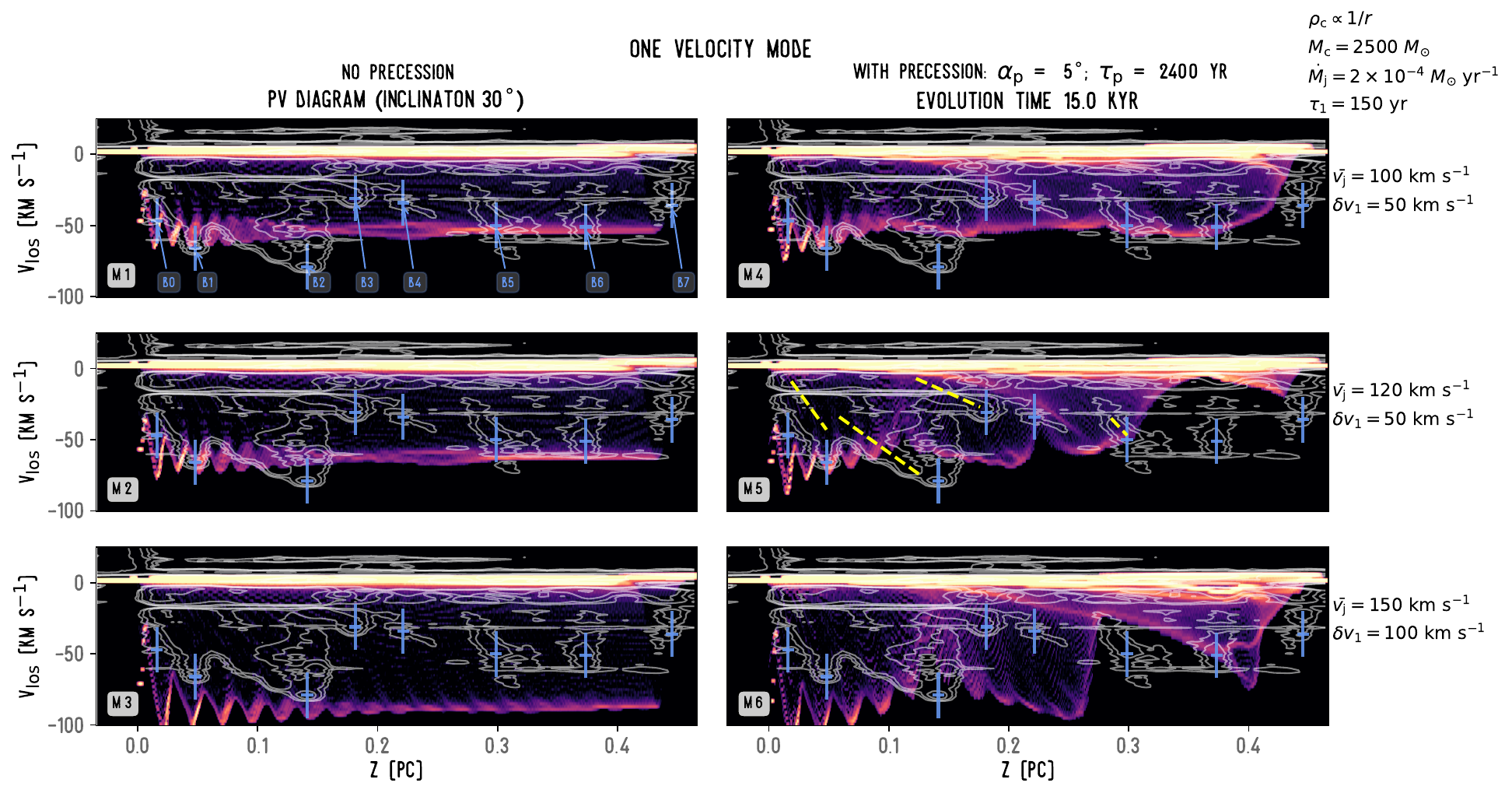}
\caption{Color map of the PV diagrams of the models with one velocity mode (M1-M6, see Table\ref{table:simulations}). We over-plotted the CO iso-contours of the PV diagram of core\#8 reported by \cite{Nony20} at 5, 10, 22 to 67 times the units of $\sigma_{\rm CO}=2.5$ mJy beam$^{-1}$. With blue crosses, we show the positions of the observed blue knots. As a reference, we draw over the PV diagram of model M5 three yellow lines which represent the ``Hubble laws'' connecting knots with low velocity structures, reported in Figure 5a of \cite{Nony20}. }
\label{fig:pv1}
\end{figure*}

\subsection{The two velocity mode case}\label{sec:two_velocity}
We added to the simulations presented in Section \ref{sec:one_velocity} a second velocity mode with a $600~\mathrm{yr}$ period, and a slightly lower amplitude than the fist mode (see Table \ref{table:simulations}). This second mode gives a separation among knots for  model without precession of $\Delta x2_\mathrm{M7} = 0.06$\,pc, $\Delta x2_\mathrm{M8} = 0.07$\,pc, and  $\Delta x2_\mathrm{M9} = 0.09$\,pc.
The second velocity mode produces internal working surfaces that would start appearing at :
\begin{equation}
    \Delta x_{ws} \approx \frac{v_1^2 \tau_2}{2 \pi \delta v_2},
\end{equation}
from the source \citep{castellanos:18}. This is strictly valid when $\delta v_2 \ll v_1$ \citep{Raga:98}). In our models M7, M8, and M9 the starting position of the working surfaces of the second mode, are  $\Delta x_\mathrm{ws,7}=0.07$, $\Delta x_\mathrm{ws,8}=0.026$, and $\Delta x_\mathrm{ws,9}=0.04$, respectively.

For the two-mode velocity case we present Figure \ref{fig:dens2}, which shows a density cut in the $XZ$-midplane at an integration time of $15$~kyr. The associated column density maps are shown in Figure \ref{fig:N2}.
The simulations without and with precession are shown in the left and right panels, respectively (see all initial parameters used in Table \ref{table:simulations}). The impact of the second mode in the velocity is clear. In this case we obtain a series of knots periodically appearing at farther distances ($\Delta x2$), which now could account for the position of the four farthest blue knots observed in core \#8.

In the simulations without precession (left panels of Figure \ref{fig:dens2}), the second velocity  mode results in a similar morphology to the case with one velocity mode, with a small variation in the distances among knots. 
Similarly to the one-velocity modes, the effect of increasing the average velocity of the jet (20 \kms  between model M7 and M8, and 30 \kms  between model M8 and M9; see left panels of Figure \ref{fig:dens2}) results in the appearance of the farthest knots at greater distances. The same behaviour is obtained for the models including precession.
However, when including the precession in the simulations (right panels of Figure \ref{fig:dens2}) the farthest knots can be better distinguished. The latter is because the material collides with the surface of the cavity carved by previous high velocity material, thus increasing the local density resulting in better defined knots.

\begin{figure*}
\epsscale{1.2}
\plotone{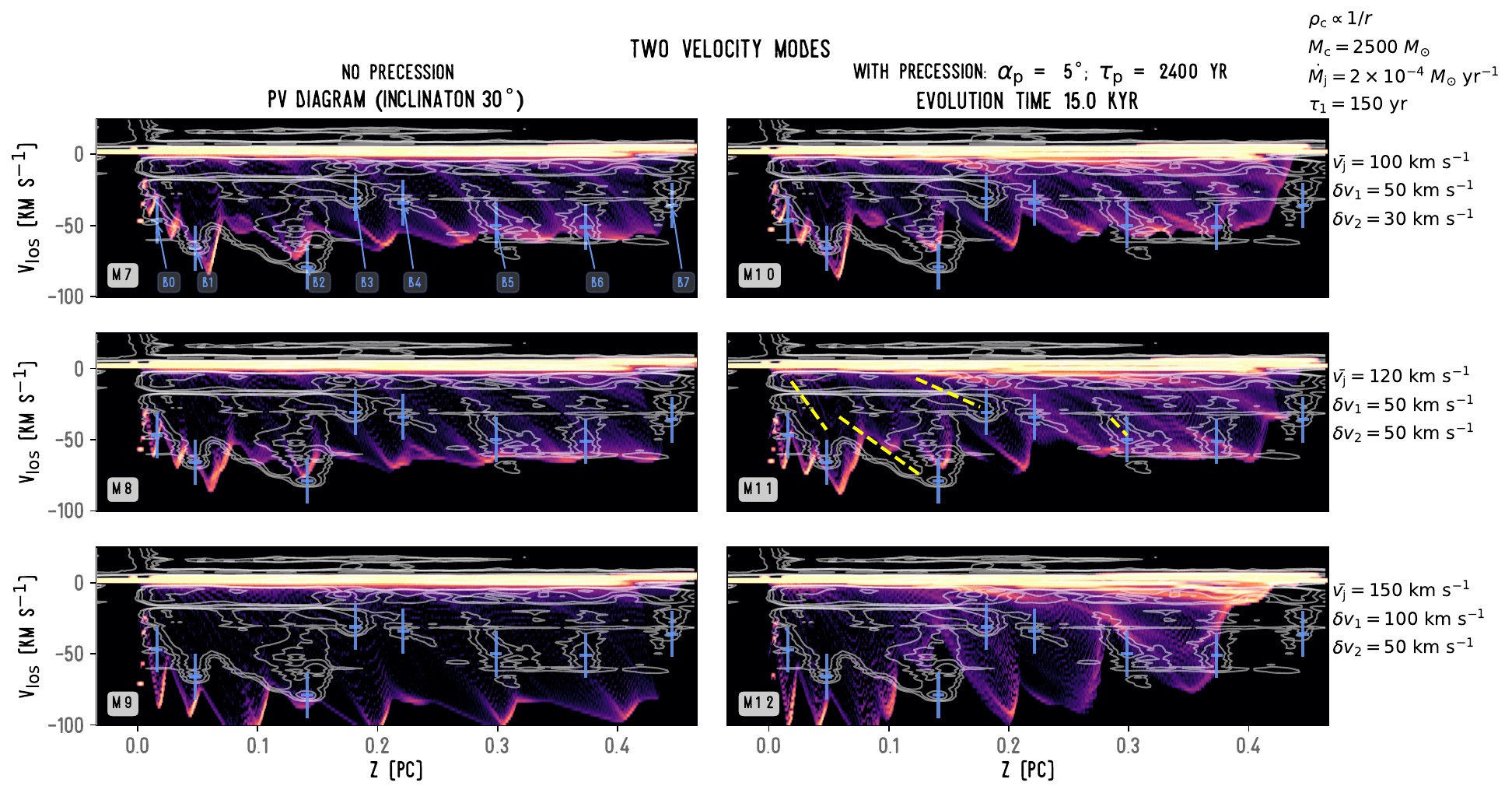}
\caption{Color map of the PV diagrams of the models with two-velocity modes (models M7-M12). Symbols have the same meaning as in Figure \ref{fig:pv1}.
\label{fig:pv2}}
\end{figure*}

\subsection{PV diagrams}\label{sec:pv}
The upper panel of Figure 5 (a) of \citet{Nony20} shows the PV diagrams along the molecular outflow of core \#8. 
In order to compare our results to the observations, 
we also created from our simulations the synthetic PV diagrams of line-of-sight (LoS) velocities as a function of projected distance (see Figures \ref{fig:pv1} and \ref{fig:pv2}). The PV diagrams were computed placing a virtual slit along the jet axis with a velocity resolution of $2~\kms$. We averaged the PV from 20 cells in the $x$-direction (perpendicular to the jet axis), resulting in an spatial resolution of $\sim 0.015$~pc, corresponding approximately to the smoothed spatial resolution in the observed PVs. 

An additional free parameter when computing PV diagrams is the relative orientation between the jet and the LoS. The PVs shown here correspond to an  angle of $60\degr$ between the jet axis and the LoS ($30\degr$ with respect to the plane of the sky). This angle is close to the most probable inclination for an outflow, and it is also consistent with the estimations for the observed outflow (see Figure~12a of \citealt{Nony20}).
With this angle, LoS velocities are reduced by a factor of two compared to the intrinsic jet velocities, while projected distances are less impacted.

We over-plotted in Figures \ref{fig:pv1} and \ref{fig:pv2} the CO contours used by \citet{Nony20}.
The PV diagrams in the one-velocity mode case without precession (Models 1, 2, and 3; left panels of Figure \ref{fig:pv1}) show the continuous chain of knots driven by $\delta v_1$. For models 1, 2, and 3, at distances $>0.18$ pc, the PV diagrams show a clear plateau at constant velocity of $\Delta V \simeq~-60 \kms$. Such plateau is observed even when including the precession (Models 4 and 5; see upper and middle right panels of Figure \ref{fig:pv1}). 
This is not in agreement with the variation of velocity observed in the data, with a velocity of knots B3 and B4 about 20~$\kms$ lower than B5 and B6 and even 40~$\kms$ lower than B1 and B2.

The PV diagrams in the two-velocity mode case without precession (Models 7, 8, and 9; left panels of Figure \ref{fig:pv2}) show clearly the knots at distances greater than $0.25$ pc, whereas the plateau observed in the analogous one-velocity mode (Models 1, 2 and 3) is not observed. 
The velocity variations observed in the models with precession (10 and 11), increasing rapidly up to $\simeq 0.12$\,pc and decreasing at larger distances, qualitatively reproduce the observations.

\begin{figure}
\epsscale{1.2}
\plotone{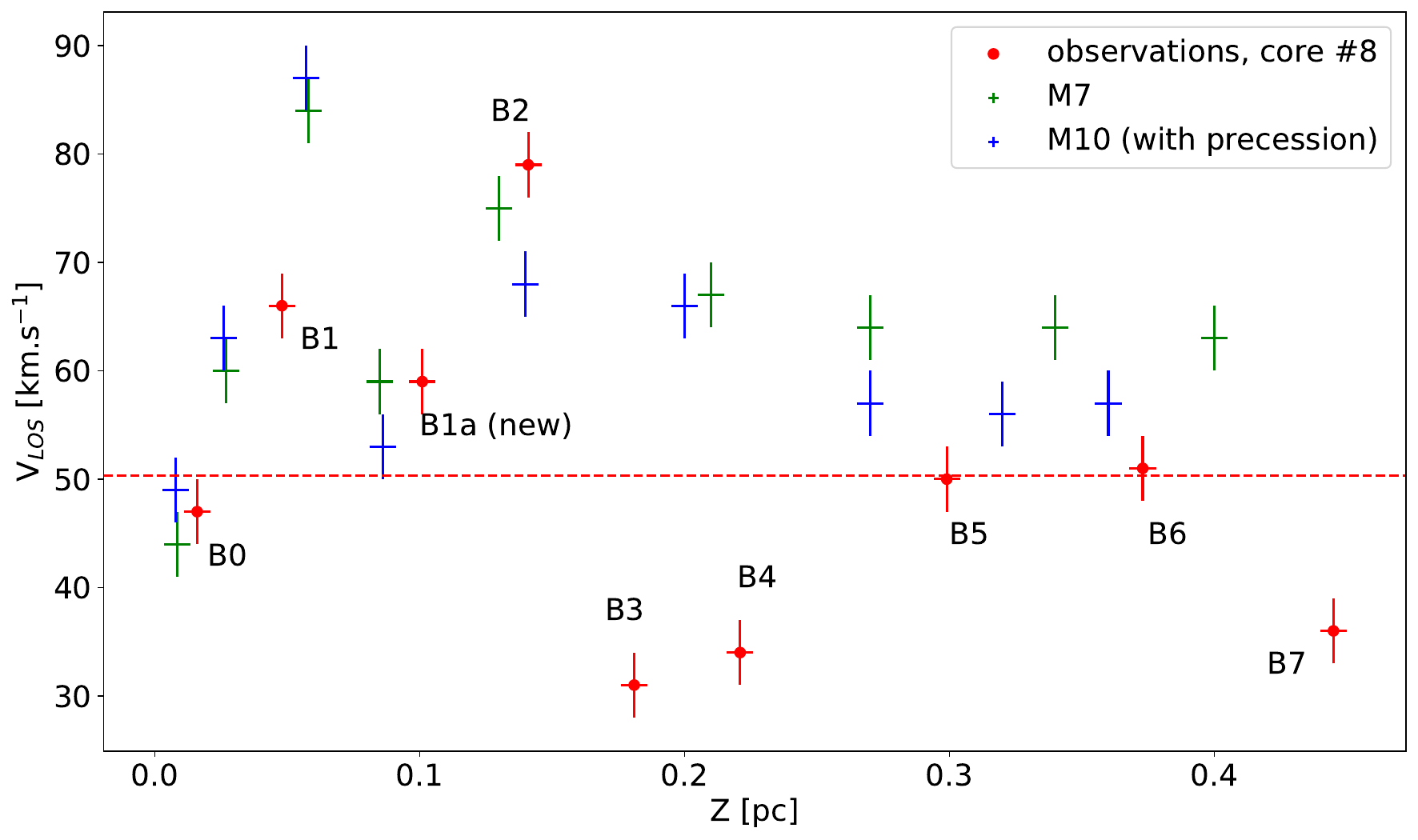}
\caption{In this Figure we show the positions and velocities of the knots identified in models M7 (green) and M10 (blue), and 
the observed knots (red) from the PV diagram of core \#8 of \cite{Nony20}. The observed knots are labeled from B0 to B6.
The error bars give the observational uncertainties in position ($5\times10^{-3}$~pc) and velocity (3 $\kms$).}
\label{fig:pv-comp}
\end{figure}

In order to quantitatively determine which of the two-velocity mode models best reproduces the observations, we measured the positions and velocities of the knots in the PV diagrams of our simulations, following the method described in \citet{Nony20}.
The average velocity of the knots is $\overline{\Delta V_{\rm knot}} \simeq 63$, $71$ and $97\,\kms$ for models M7-M10, M8-M11 and M9-M12, respectively. These values are about 60\% of the average injection velocity.
Models M7 and M10 have an average knot-velocity similar to that reported by \cite{Nony20} for core \#8 ($\overline{\Delta V_{\rm knot}}=\,50\,\kms$), and therefore our best model candidates.

Interestingly, based on the model predictions for the knots velocity and position, we found between knots B1 and B2 of the observed PV diagram (Figure 5 (a) of \citealt{Nony20}) an over-density which was not previously reported. Although this structure does not correspond to all our criteria for defining a knot -- an over-density at a local velocity maximum --, it could arguably correspond to an independent ejecta. We labeled this structure B1a and report its coordinates in Table~\ref{table:knots}.

In Figure~\ref{fig:pv-comp} we compare the relative positions and velocities of the observed knots of core \#8 outflow (B0 to B7) with that of the knots found in models M7 and M10.
We associated each of the observed knots to its closest counterpart in the models, and measured the relative velocity and position offsets.
We first note that the inclusion of precession in model M10 compared to model M7 introduces only a small difference in the positions and velocities of the knots (up to 8$\,\kms$ and $0.04$ pc, respectively). Model M10, which includes precession, reproduces slightly better the observations than model M7. The average position and velocity differences are indeed smaller: $0.018$ pc vs $0.025$ pc, and $11.5$ vs $13.1\,\kms$, for models M10 and M7 respectively.
Model M10 performs specially better in reproducing the knots at large distances (B5 and B6). However, neither model reproduces the velocity of knots B3 and B4 and, conversely, both models predict a knot at high velocity (86\,$\kms$ at 0.06 pc) which does not have an observational counterpart.

\section{Conclusions and discussion} \label{sec:conclusions}
Jets and outflows in star forming regions often present substructure in the form of knots, which indicates that the material ejection could be episodic \citep[e.g.,][]{Qiu19, Li2020}. Significant ejection variability has been observed in the high-mass protocluster W43-MM1 \citep{Nony20}, with the outflow of core \#8 presenting the most numerous knots and therefore constituting the target of our study.

We ran 3D hydro-dynamical simulations of a proto-stellar jet and studied the effect of including a periodic velocity ejection variation, with one and two velocity modes. The values of the velocity-modes  were selected to match the observed periodicity of core \#8 reported in \cite{Nony20}. Additionally, we explored the effect of the precession of the jet.

We ended up with 12 different models (see Table \ref{table:simulations}), for which we studied the knots formed in the simulations making use of density (and column density) maps, as well and PV diagrams  (see Figures \ref{fig:dens1}-\ref{fig:pv2}).\\

We summarize our results as follows:

\begin{itemize}
    \item Models with one-velocity ejection modes (M1 to M6) do not match the observed velocity variations of core \#8. Instead, models with 
    two-velocity modes (M7 to M12) have a better agreement with the observed velocity and positions of the knots observed in core \#8. 

    \item The observed outflow shows evidence of precession. Precession leads to periodic oscillations of the outflow direction, and it also has an impact on the velocity of the knots (see, e.g., the comparison between M7 and M10). 

    \item Model M10, with two-velocity modes, precession, and an average jet injection velocity $v_1 = 100\,\kms$ reproduces fairly well the observations of core \#8 even without fine tuning. Very interestingly, our simulations predict the position and velocity of a new knot (B1a, see Figure \ref{fig:pv-comp}) that was not previously reported in \cite{Nony20}.
\end{itemize}

Both the variability modes of the ejection and the precession of the jet are likely associated to mechanisms occurring at the protostellar disk scale ($\sim 100$~au). Perturbations in a binary system  have been proposed as a possible phenomena at play \citep{Terquem99,Raga09}.

The physical processes in the molecular outflow of core MM1\#8 could be different from those of the intermediate-mass Cepheus E core, which was recently studied with CO observations and numerical simulations  \citep{Schutzer22,Rivera23}. \cite{Rivera23} showed that one-velocity ejection variations successfully reproduce the observed features, and especially the bimodal distribution of the dynamical timescales of the knots. On the contrary, we showed in this work that a two-velocity ejection mode is necessary to reproduce the observed velocity variations among the knots. The differences might come from the different environment where these objects are, or the existence of a companion. 

In general, the morphological differences in distinct outflows might come from the different physical mechanisms acting in the disk. For example, thermal instabilities and the existence of a companion, might give rise to quasi-periodic 
variations in the accretion rate, which might translate in a periodic difference in the positions of the knots in an outflow, depending on the outburst periodicity \citep{Audard14}. Morphological differences might also come from the evolutionary stage of the outflows, their inclination, and precession, which might translate into shorter or longer outflows. In addition, the environment density where the outflow propagates, could also shape its morphology.

Our modelling of the outflow from core \#8 is a step forward to a better understanding of intermediate-mass outflows. This intermediate-mass object (18M$_{\odot}$) indeed has the potential to form a star of 6-9M$_{\odot}$ (when assuming 1/3 to 1/2 efficiency). Moreover, it lies within a cluster of outflow driven by low- to high-mass cores \citep{Nony20}. 
Then, we are studying an object in the border between an intermediate and a high-mass protostars. 
Our work thus demonstrates that the two-mode velocity variability, initially proposed for optical jets from low-mass protostars, could also explain the morphology of a molecular outflow from an intermediate mass core.
Further simulations of protostellar outflows should be carried to explore the outflows temporal variation and the merging of knots. 
\\
\\
\\

\section*{Acknowledgments}

We would like to thank the referee for giving very constructive comments and suggestions, which widely help to improve this paper.
VL acknowledges the support of CONAHCyT. VL wishes to thank A. Raga for discussions and suggestions throughout this work, and the life sailed through the years.
VL and AE acknowledge support from PAPIIT-UNAM grant IN113522.
TN and RGM acknowledge support from UNAM-PAPIIT project IN108822 and from CONACyT Ciencia de Frontera project ID 86372. TN also acknowledges support from the postdoctoral fellowship program of the UNAM.


\newpage

\appendix
\counterwithin{figure}{section}

\section{Column density maps}
\label{sec:app}
In this appendix we present the column density maps of the simulations presented in the main body of this article. 

\begin{figure}[htp]
    \epsscale{0.95}
    \plotone{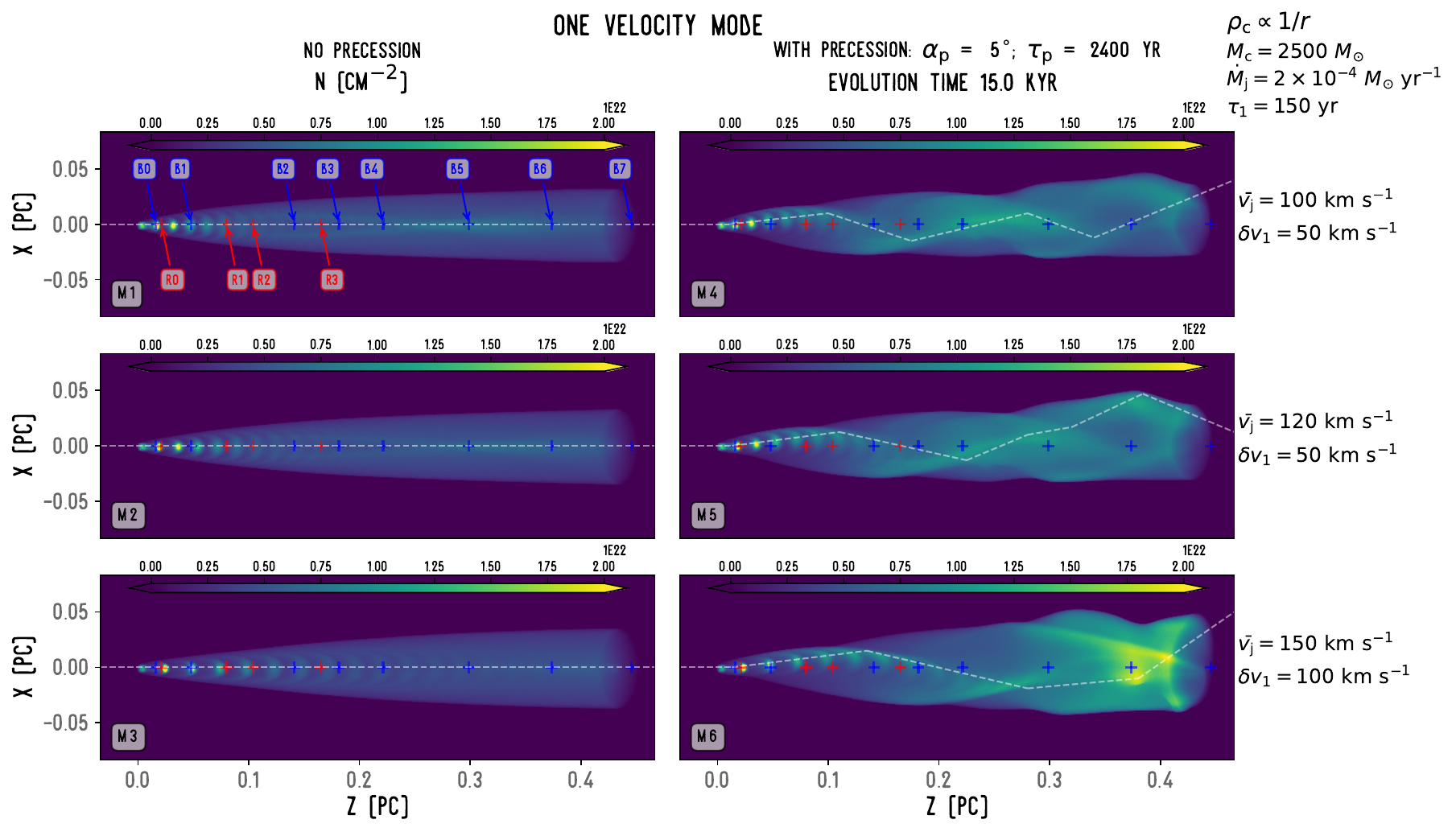}
    \caption{Column density in the same projection as the PV figures for the models with one-velocity mode. The white dashed line denotes the position of the virtual slits used to produce the PV diagrams.}
    \label{fig:N1}
\end{figure}

\begin{figure}[htp]
    \epsscale{0.95}
    \plotone{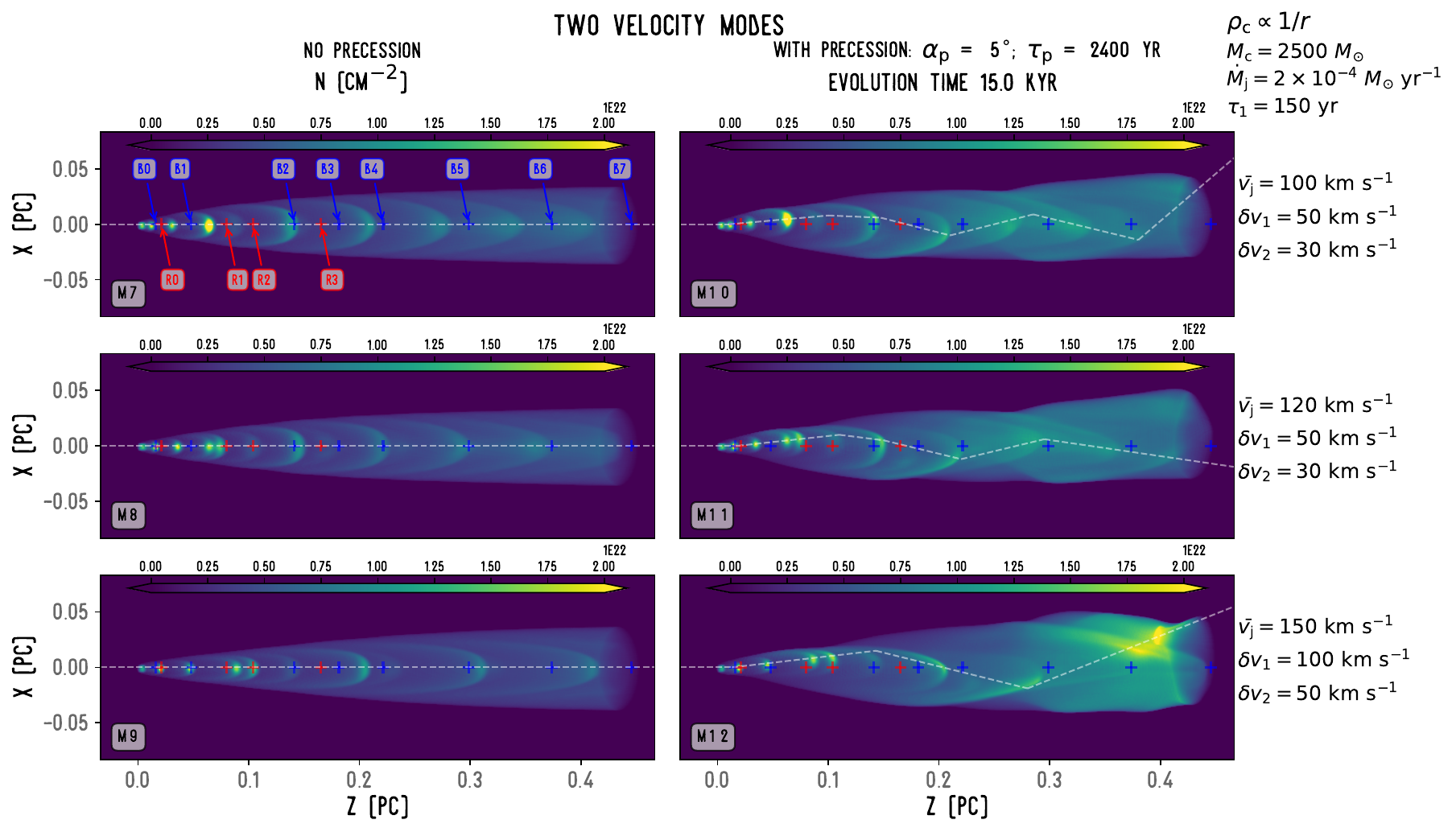}
    \caption{Same as Fig. \ref{fig:N1}, for the models with two-velocity modes.}
\label{fig:N2}
\end{figure}

Figure \ref{fig:N1} shows the column density with the same projection as the PV diagrams for the models with one velocity mode (see Figure \ref{fig:pv1}). Figure \ref{fig:N2} shows the column density with the same projection as the PV diagrams for the models with two velocity modes (see Figure \ref{fig:pv2}).

In both Figures \ref{fig:N1} and  \ref{fig:N2} the left panels show the models without precession, and the right panels show the models with precession (with a half opening angle of $5\degr$ and a precession period of $2400~\mathrm{yr}$). The rows of Figures \ref{fig:N1} and   \ref{fig:N2} show the models with different velocity variability, different average velocity of the jet, and different amplitude of the fluctuation of the velocity. The blue and the red crosses mark the position of the blue- and red-shifted knots for core \#8 (see Table~\ref{table:knots}).

\newpage
\bibliography{ms}{}
\bibliographystyle{aasjournal}

\end{document}